\documentclass{llncs}
\usepackage{llncsdoc}

\usepackage[linesnumbered,ruled,vlined]{algorithm2e}
\usepackage{mathtools}
\usepackage[space]{cite}
\usepackage{url}

\usepackage{subfig}
\usepackage{graphicx}
\usepackage{booktabs}
\usepackage{subcaption}
\usepackage{caption}

\usepackage{algorithmic}

\usepackage{amssymb}
\usepackage{slashbox}
\usepackage{float}
\usepackage{listings}
\usepackage{amsmath}
\usepackage{rotating}
\usepackage{qcircuit}
\usepackage{graphics}
\usepackage{braket}

\usepackage{array}
\usepackage{enumitem}
\usepackage{arydshln}

\usepackage{makecell} % makecell 패키지 사용 - Xhline을 위한
\newcommand{\mc}[1]{\mathcal{#1}}
\usepackage{kotex}
\usepackage{xcolor}
\usepackage{tikz}
\usetikzlibrary{arrows.meta, positioning}
\usepackage{adjustbox}
\usepackage{makecell}

\title{Partial Blind Quantum Computation: A Framework for Selective Circuit Protection}
\author{
Youngkyung Lee\inst{1} \and
Doyoung Chung\inst{1} \thanks{Corresponding author}
}
\authorrunning{Y. Lee and D. Chung.}
\institute{
Information Security Research Division, ETRI, Daejeon, 34129, Korea \inst{1} \\
\email{\{youngklee, thisisdoyoung\}@etri.re.kr}
}

\begin{document}
\maketitle

% \keywords{Blind Quantum Computing (BQC) \ quantum cryptanalysis \ quantum circuit \ quantum homomorphic encryption}
\begin{abstract}
Quantum computing is rapidly advancing toward cloud-based services, raising significant concerns about the privacy and security of computations outsourced to untrusted quantum servers. Universal Blind Quantum Computation (UBQC) protocols enable clients with limited quantum resources to delegate computations while concealing both inputs and circuit details. However, applying UBQC uniformly to an entire quantum circuit incurs additional quantum resources and computational overhead, which can be a significant burden in practical implementations. In many cases, such as Grover’s algorithm, only specific subroutines—like oracles—contain sensitive information, while the rest of the circuit does not require the same level of protection. Therefore, selectively applying UBQC to critical components can enhance computational efficiency while maintaining security.

In this work, we propose a selective application of UBQC that targets only the critical components of quantum circuits. By integrating techniques from Quantum Homomorphic Encryption (QHE) and UBQC, our approach secures the sensitive subcircuits while allowing the remaining, non-sensitive portions to be executed more efficiently. In our framework, UBQC-protected sections output quantum states that are encrypted via bit-flip and phase-flip operations, and we devise a mechanism based on selective X and Z gate corrections to seamlessly interface these with unprotected sections. We provide a security analysis demonstrating that our selective UBQC approach preserves universality, correctness, and blindness, and we illustrate its practical advantages through an application to Grover’s algorithm. This work paves the way for more efficient and practical secure quantum computing on near-term devices.

\end{abstract}


\section{Introduction}

Quantum computing has emerged as a transformative technology capable of addressing computational problems intractable for classical systems.
It has shown significant promise in areas such as cryptography, combinatorial optimization, and quantum simulation \cite{AJL06, Preskill2018, Arute2019, Park2024}.
As the physical realization of quantum devices remains complex and costly, major quantum hardware providers have begun offering access to quantum processors via remote platforms and APIs\cite{IBM2020, GoogleQuantumAI}, making remote access to quantum resources feasible without dedicated infrastructure.

However, this cloud-based computing model introduces a critical challenge: users must trust external quantum servers with computations that may involve highly sensitive quantum data or proprietary algorithms.
The risk of information leakage arises naturally when computations are outsourced to infrastructure not controlled by the user \cite{Mantri2017}.
In response, the quantum cryptography community has explored secure delegation protocols, such as blind quantum computation and quantum homomorphic encryption, that aim to preserve data privacy in untrusted environments \cite{Childs2005, Arrighi2006, Broadbent2015Delegating}.

Universal Blind Quantum Computation (UBQC; referred to hereafter as UBQC or simply BQC) is one of the most prominent cryptographic protocols proposed to address these delegation-related privacy concerns. It enables a client with limited quantum capabilities to delegate quantum computations to an untrusted server while keeping both the data and the algorithm confidential \cite{broadbent2009universal, fitzsimons2017Priv}. Over time, subsequent work has reduced the quantum requirements on the client side \cite{morimae2013blind, morimae2016measurement} and introduced verifiability features to detect server misbehavior \cite{fitzsimons2017uncond, aharonov2017, gheorghiu2019}.

Despite these advances, applying BQC uniformly across an entire quantum circuit results in substantial computational overhead due to the need for quantum state preparation and classical communications between server and client.
In many practical scenarios, only specific subroutines, such as oracle calls in Grover’s algorithm, contain sensitive information, while the remaining parts of the circuit can be executed without confidentiality concerns.
This observation motivates the development of more efficient schemes that selectively apply BQC only where privacy is necessary.

Building on this observation, we introduce the notion of Partial Blind Quantum Computation (PBQC) to describe protocols in which blindness is selectively applied only to subcircuits that handle sensitive information, rather than enforced across the entire computation.
This selective approach reflects both the scope of privacy protection and the modular structure of practical quantum algorithms such as the clear separation of oracle and diffusion operations in Grover’s algorithm.
By focusing computational overhead only where it is necessary, PBQC offers a more efficient and flexible framework for secure delegated quantum computation.
In this work, we develop a practical PBQC framework that integrates the principles of blind quantum computation and circuit partitioning, and we demonstrate its applicability through the selective blinding of oracle components in Grover’s algorithm.

\subsection{Background}

Applying UBQC to an entire quantum circuit is often prohibitively resource-intensive due to its interactive nature and communication overhead.
To address this, we adopt a hybrid approach in which BQC is selectively applied only to those parts of the circuit where both data and circuit structure must remain private.
For the remaining parts, where circuit confidentiality is unnecessary, we utilize Quantum Homomorphic Encryption (QHE)~\cite{fisher2014quantum} solely to protect the quantum data.

Quantum computation operates under fundamentally different principles, and algorithms such as Grover’s and Shor’s typically initialize inputs as $\ket{0}$ states, encoding logical information directly into the circuit, often through oracle functions.
This tight coupling between data and circuit design raises stricter privacy concerns, requiring notions such as \textit{circuit blindness}, where no information about the structure of the circuit is leaked to the server.

UBQC was introduced to address these privacy needs. Since the original BFK09 protocol \cite{broadbent2009universal}, considerable research has focused on achieving secure delegation with minimal quantum capabilities on the client side. These protocols~\cite{broadbent2009universal, morimae2013blind, fitzsimons2017uncond} aim to ensure privacy of both input and computation, even in the presence of a malicious server.

UBQC is built upon the Measurement-Based Quantum Computation (MBQC) model~\cite{raussendorf2001}, where each quantum gate can be decomposed into a sequence of measurements of the form $J(\theta) = HR(\theta)$.
In this framework, the server performs measurements at angles provided by the client, who randomizes these angles to conceal the underlying computation.
Although MBQC is central to UBQC, most current quantum devices are based on the gate model and are expected to translate gate-based circuits into MBQC form, at least in the near term.
In doing so, UBQC incurs a non-negligible overhead of $O(n)$. Given the limited quantum resources available for processing on quantum computers, this increase cannot be disregarded.

In practice, however, the portions of a quantum circuit that require full UBQC protection are often limited. For instance, in Grover’s algorithm, if revealing the use of the algorithm is not a concern, only the oracle function—which typically encodes the sensitive component—needs to be kept private. Furthermore, even within the oracle, the sensitive content may be confined to a specific function $f$ applied to a subset of the input space.

Motivated by this observation, we propose a selective delegation model referred to as PBQC, where BQC is applied exclusively to sensitive subcircuits.
The remaining parts are executed using QHE, which ensures the confidentiality of quantum data while permitting the server to process the (non-sensitive) circuit openly.
Unlike BQC, QHE does not offer circuit blindness; instead, it serves as a lightweight mechanism to protect the client’s input and output states.

A key technical challenge in PBQC is achieving seamless integration of BQC-protected outputs, which are typically randomized through bit-flip and phase-flip encryptions, into QHE-based execution regions in a seamless manner.
We address this challenge by introducing a bridging mechanism that preserves the blindness of the protected subcircuits and maintains end-to-end data confidentiality across the entire computation.
This hybrid strategy enables efficient and privacy-preserving delegated quantum computation by combining the strengths of BQC and QHE according to the security requirements of each circuit component.

\subsection{Related Work}

Since UBQC was first introduced by Broadbent, Fitzsimons, and Kashefi in 2009 \cite{broadbent2009universal}, significant research has focused on reducing the quantum requirements on the client side.
Morimae and Fujii proposed a protocol where the client performs only single-qubit measurements \cite{morimae2013blind}, while verifiable BQC protocols have been developed to ensure correctness against a potentially dishonest server \cite{fitzsimons2017uncond}.
Variants tailored for near-term devices have also emerged—for example, Shingu et al. \cite{shingu2021variational} proposed a secure delegation scheme for variational quantum algorithms on NISQ hardware.
Although recent advances suggest that fully classical clients are theoretically possible \cite{mahadev2018a, mahadev2018b, cojocaru2019}, such protocols impose substantial overhead on the server and remain impractical for near-term deployment.
Recent studies have further emphasized efficiency. Zhang proposed a succinct BQC protocol using a random oracle, enabling the client to prepare a fixed number of quantum states independent of circuit size \cite{zhang2020succinct}. In parallel, Cao et al. introduced a multi-agent BQC scheme that avoids the need for universal cluster states by distributing the computation across multiple entangled quantum agents \cite{cao2023multiagent}, improving scalability and reducing complexity.

In parallel, QHE has emerged as a non-interactive alternative for secure delegated quantum computation. Early experimental demonstrations \cite{zeuner2021} showed that encrypted quantum states can be processed on photonic hardware without revealing data to the server. Recent protocols aim to reduce interaction overhead by introducing two-round QHE schemes \cite{shang2023} or dynamic correction methods for universal quantum circuits \cite{chang2023}, striking a balance between communication efficiency and computational generality. Meanwhile, verifiable QHE frameworks \cite{he2024} and formal analyses of privacy trade-offs \cite{hu2023} have enhanced the theoretical foundation of encrypted quantum computation. Furthermore, distributed and multi-party extensions of QHE \cite{chen2023, pan2024} have been proposed to support scalable secure computing across quantum networks.

Recently, Selectively Blind Quantum Computation (SBQC) \cite{poshtvan2025sbqc} was proposed as a relaxation of UBQC, enabling the client to hide only the choice among a known set of computations. While SBQC and our work both aim to balance information leakage and communication efficiency in secure quantum delegation, they pursue this trade-off through fundamentally different approaches. SBQC reduces quantum communication by masking only the differences between candidate computations, while preserving information-theoretic security, and establishes no-go results showing that no server-side quantum process can replicate or amplify encrypted states without compromising blindness.

In contrast, our approach introduces a hybrid model that selectively applies BQC only to sensitive subcircuits within a larger computation. This allows the client to delegate computations securely while minimizing quantum resource overhead. Our method also incorporates QHE-inspired techniques to manage encrypted intermediate states flowing between protected and unprotected regions, enabling fine-grained control over which components require blindness. By targeting partial blindness at the subcircuit level, we provide a flexible and resource-efficient alternative to fully blind delegation.

\subsection{Our Contributions}

This work introduces Partial Blind Quantum Computation (PBQC), a hybrid delegation framework that applies blindness selectively to privacy-sensitive subcircuits. Our main contributions are as follows:

\begin{itemize}
\item \textbf{PBQC Framework Design:} We propose a general framework for selectively applying UBQC only to subcircuits requiring confidentiality, while executing the remaining portions under QHE. This hybrid design reduces resource overhead without compromising circuit correctness or privacy (Section 3).

\item \textbf{QHE-Compatible UBQC Protocol:} We develop a modified UBQC protocol that accepts QHE-encrypted inputs and produces QHE-encrypted outputs, ensuring seamless integration between blinded and unblinded regions of the circuit. This construction enables circuit-wide end-to-end confidentiality while supporting modular composition of BQC and QHE subcomponents (Section 4).

\item \textbf{Implementation and Evaluation on Grover’s Algorithm:} We demonstrate the practicality of PBQC by applying it to a 2-qubit Grover algorithm.
    The oracle is protected via the proposed QHE-compatible UBQC protocol, while the initialization and diffusion operations are evaluated under QHE.
    A full simulation using Qiskit, an open-source quantum computing framework developed by IBM, confirms both correctness and resource efficiency, showing reduced qubit usage and measurement depth compared to standard UBQC (Section 5).

\end{itemize}

Furthermore, when BQC‑protected segments combine with non‑BQC segments, they form a complementary resource relationship. The MBQC nature of a BQC‑protected section uses a large cluster of qubits, but once each qubit is measured and its entanglement consumed, that physical qubit becomes free. Quantum circuits that realize the same logical functionality can often be implemented in alternative ways that trade qubit count against circuit depth.
Depth‑optimized implementations, however, often call for extra ancilla qubits. Prior studies~\cite{Selinger2013, DWHW2020, DCKF2023} show that this demand can be met while incurring minimal additional cost by resetting qubits whose roles have concluded and reusing them. Similarly, our approach resets the qubits released by the preceding BQC block and promptly reuses them as ancillae for the subsequent non‑BQC section. Because the gate‑level layout of the non‑BQC portion is largely preserved (except for non‑Clifford gates), we can choose depth‑efficient implementations without raising the peak qubit count, thereby further enhancing the circuit’s overall resource efficiency.

\section{Preliminaries}
\label{sec:preliminaries}

The following preliminaries summarize core concepts relevant to our work.
First, we explain the principle of UBQC, which enables a client to delegate quantum computation to a server while preserving the privacy of the client’s data and operations. We then outline the basics of QHE, which allows computations on encrypted quantum data, ensuring privacy throughout the process.

\subsection{Universal Blind Quantum Computation (UBQC)}

The core of UBQC lies in the structure of MBQC~\cite{raussendorf2001}, where quantum gates are implemented through adaptive measurements on entangled resource states. Fig.~\ref{fig:input_teleportation} illustrates the basic teleportation primitive in MBQC: a qubit $\ket{\psi}$ entangled via a $CZ$ gate with an ancillary $\ket{+}$ qubit can be teleported onto the ancillary qubit by measuring the original qubit in the $X$-basis. The resulting state is $X^m H \ket{\psi}$, where $m \in \{0,1\}$ is the measurement outcome.

\begin{figure}
\centering
\makebox[\textwidth]{
    \Qcircuit @C=1em @R=.7em {
    \lstick{|\psi\rangle} & \ctrl{1} & \gate{H} & \meter & \cw & \rstick{m} \\
    \lstick{|+\rangle}    & \ctrl{-1}& \qw      & \qw    & \qw & \rstick{X^{m}H|\psi\rangle} \\
    }
}
\caption{Basic state teleportation used in MBQC.}
\label{fig:input_teleportation}
\end{figure}

A more general form involves measuring in the $\ket{+_\theta}$ basis, implementing the gate $J(\theta) = H R_Z(\theta)$ on the teleported state. Fig.~\ref{fig:j_basis_measurement} illustrates the gate teleportation primitive in MBQC. The resulting state is $X^m J(-\theta) \ket{\psi}$, where $m \in \{0,1\}$ is the measurement outcome.
Since all single-qubit unitaries can be decomposed into $J(\theta)$ gates, this property enables universal computation over cluster states.

In UBQC, the client leverages this property to encode the circuit logic within measurement instructions, while hiding it through random rotations and classical obfuscation.

\begin{figure}[h]
\centering
\makebox[\textwidth]{
    \Qcircuit @C=1em @R=.7em {
    \lstick{\ket{\psi}} & \ctrl{1} & \gate{R_Z(-\theta)} & \gate{H^{\dagger}} & \meter & \cw & \rstick{m} \\
    \lstick{\ket{+}}    & \ctrl{-1}& \qw               & \qw      & \qw    & \qw & \rstick{X^{m}HR_Z({-\theta})|\psi\rangle} \\
    }
}
\caption{
Gate teleportation of $J(\theta) = H R_Z(\theta) $ via rotated-X-basis measurement in MBQC.}
\label{fig:j_basis_measurement}
\end{figure}

\paragraph{Blinded Measurement-Based Computation.}

Computation proceeds through a series of adaptive single-qubit measurements on an entangled graph state.
To initialize the protocol, the client prepares and sends single-qubit states of the form:
\[
\ket{+_{\theta}} = \frac{1}{\sqrt{2}}(\ket{0} + e^{i\theta} \ket{1}), \quad \theta \in \{0, \pi/4, \dots, 7\pi/4\},
\]
where each $\theta$ is chosen uniformly at random to blind the true computation angles.
These qubits are embedded into a fixed entangled structure called the *brickwork state*, a universal resource for MBQC.
The entanglement operations to create the brickwork state are performed entirely by the server.

The client then sends encrypted measurement angles:
\[
\delta = \phi + \theta + \pi r,
\]
where \(\phi\) is the desired measurement angle for the computation, \(\theta\) is the blinding phase used during state preparation, and \(r \in \{0,1\}\) is a random bit to further mask the outcome.
The server performs the measurement in the basis \(\{\ket{+_{\delta}}, \ket{-_{\delta}}\}\), and returns the classical result to the client, who adjusts the subsequent measurement instructions accordingly.

\paragraph{Security Properties.}

The UBQC protocol satisfies the following cryptographic guarantees~\cite{broadbent2009universal}:
\begin{itemize}
  \item \textbf{Correctness:} The output is correct modulo known Pauli corrections, which the client can track and reverse.
  \item \textbf{Blindness:} The server learns nothing about the input, output, or the computation itself; only the size of the circuit is revealed.
  \item \textbf{Universality:} Any quantum circuit can be expressed as a measurement pattern on the brickwork state.
\end{itemize}

\subsection{Quantum Homomorphic Encryption (QHE)}
\label{sec:preliminaries-QHE}

\paragraph{Homomorphic Encryption} is a cryptographic technique that enables computation directly on encrypted data without revealing the underlying plaintext. In a fully homomorphic encryption (FHE) scheme, three core algorithms are defined: $\mathsf{Encrypt}(m)$ to encrypt a message $m$, $\mathsf{Eval}(f, \mathsf{Encrypt}(m))$ to homomorphically apply a function $f$ on the ciphertext, and $\mathsf{Decrypt}(\cdot)$ to recover the result $f(m)$.
The function $f$ is not directly applied to the message $m$, but homomorphically evaluated on its encryption, producing an encrypted output corresponding to $f(m)$~\cite{gentry2009, boyle2018circuit}.

\paragraph{Quantum Homomorphic Encryption} is the quantum analogue of FHE, extending the same security goal to the quantum setting—namely, enabling quantum computations on encrypted quantum data without revealing any information about the quantum data~\cite{fisher2014quantum, dulek2020privacy}.
QHE typically builds on the Quantum One-Time Pad (QOTP)~\cite{ambainis2000, boykin2003}, which encrypts a quantum state $\ket{\psi}$ by applying random Pauli operators $X^a Z^b$, with secret classical bits $a, b \in \{0,1\}$ as the encryption key.

Let $\ket{\psi} = \alpha \ket{0} + \beta \ket{1}$ be the plaintext state. The encrypted state is:
\[
\ket{\psi_{\text{enc}}} = X^a Z^b \ket{\psi}.
\]

Homomorphic evaluation proceeds by transforming this encrypted state according to a desired quantum circuit $C$. The evaluation strategy and the corresponding key update rules depend on whether the gates in $C$ belong to the Clifford group or not.

\paragraph{Clifford Gates.}

For Clifford gates (e.g., $H$, $S$, $CNOT$), the transformation of Pauli-encrypted states can be tracked classically. This allows a QHE evaluator to apply gates directly to the encrypted qubits while the client updates the QOTP keys accordingly.

Let $U$ be a Clifford operator. Then:
\[
U X^a Z^b \ket{\psi} = X^{a'} Z^{b'} U \ket{\psi},
\]
for some new keys $a', b'$ that can be computed classically using Pauli conjugation rules. Table~\ref{tab:clifford-update} summarizes the key update rules.

\begin{table}[h]
\centering
%\caption{QOTP key update rules for Clifford gates.}
\caption{
Key update rules for quantum one-time pad (QOTP) encryption under Clifford gate evaluation.
Each gate transforms the classical encryption keys $(a,b)$ applied to a qubit encrypted as $X^a Z^b \ket{\psi}$.
For two-qubit gates like CNOT, the update is applied pairwise across the control and target qubits.
}
\vspace{0.2cm}
\label{tab:clifford-update}
\begin{tabular}{|c@{\hspace{1em}}|c@{\hspace{1em}}|c|}
\hline
\textbf{Gate} & \textbf{Input Key $(a,b)$} & \textbf{Output Key $(a',b')$} \\
\hline
$X$ & $(a, b)$ & $(a, b)$ \\
$Z$ & $(a, b)$ & $(a, b)$ \\
$H$ & $(a, b)$ & $(b, a)$ \\
$S$ & $(a, b)$ & $(a, a \oplus b)$ \\
$CNOT$ & $(a_1, b_1), (a_2, b_2)$ & $(a_1, b_1 \oplus b_2), (a_1 \oplus a_2, b_2)$ \\
\hline
\end{tabular}
\end{table}

These update rules enable fully homomorphic evaluation over Clifford circuits without any quantum interaction between client and server.\\

\paragraph{Non-Clifford Gates.}
Non-Clifford gates, such as the $T$-gate, do not preserve the structure of QOTP encryption under conjugation. As a result, their homomorphic evaluation necessitates an auxiliary mechanism that combines quantum and classical interaction.

A representative method, originally proposed by Fisher \textit{et al.}~\cite{fisher2014quantum}, employs a gadget-based protocol in which the client prepares and transmits an encrypted auxiliary state $S^y Z^d \ket{+}$ alongside the QOTP-encrypted data qubit $X^a Z^b \ket{\psi}$. The server then proceeds with three steps:
(i) it entangles the data and auxiliary qubits using a $CNOT$ gate,
(ii) applies the non-Clifford gate $T$ to the data qubit, and
(iii) performs a measurement, yielding a single classical bit $c$.

Prior to this measurement, the client computes a classical control bit
\[
x = a \oplus y
\]
and transmits it to the server, who uses it to apply a conditional $S^x$ gate on the auxiliary qubit. After receiving the measurement outcome $c$, the client performs a key update using the following rule~\cite{fisher2014quantum}:
\[
a'' = a \oplus c, \qquad
b'' = a(c \oplus y \oplus 1) \oplus b \oplus d \oplus y.
\]

This single round of classical interaction---sending \(x\) and receiving \(c\)---suffices to evaluate the non-Clifford gate homomorphically, without leaking any information about the underlying quantum data.
The full protocol, including its division into three logical stages—client-side preparation, server-side gate evaluation, and client-side key update—is illustrated in Fig.~\ref{fig:non-clifford-qhe}.

\begin{figure}[h]
\centering
\[
\Qcircuit @C=1em @R=1em {
\lstick{\ket{\psi}} & \gate{X^a} & \gate{Z^b} & \qw  & \push{\rule{1em}{0pt}} & \gate{T} & \targ & \meter & \cw & \push{\rule{1em}{0pt}} & \cw & \cw & \cw & \rstick{c} \\
\lstick{\ket{+}}    & \gate{S^y} & \gate{Z^d} & \qw  & \push{\rule{1em}{0pt}} & \qw      & \ctrl{-1} & \gate{S^x} & \qw & \push{\rule{1em}{0pt}} &  \gate{X^{a''}} & \gate{Z^{b''}} & \qw & \rstick{T\ket{\psi}}\\
\lstick{x = a \oplus y} & \cw & \cw & \cw & \push{\rule{1em}{0pt}} & \cw & \cw & \cw\cwx
}
\]
\vspace{-1em}
\begin{center}
% 위 3구역 라벨
\begin{minipage}[t]{0.32\textwidth}
\centering
{Client-side encryption and auxiliary state preparation}
\end{minipage}
\hfill
\begin{minipage}[t]{0.32\textwidth}
\centering
{Server-side evaluation using client-supplied control bit \(x = a \oplus y\)}
\end{minipage}
\hfill
\begin{minipage}[t]{0.32\textwidth}
\centering
{Client-side final key update and decryption}
\end{minipage}
\end{center}

\caption{Evaluation of a non-Clifford gate $T$ on QHE-encrypted input.
The circuit is divided into three logical stages:
(1) encryption of the data qubit and preparation of an auxiliary state with client keys \((a,b)\) and \((y,d)\);
(2) server-side execution of the \(T\) gate using the client-supplied control bit \(x=a\oplus y\);
and (3) client-side decryption and key update, yielding
\(a''=a\oplus c\) and
\(b''=a(c\oplus y\oplus1)\oplus b\oplus d\oplus y\).
This protocol follows Fisher \emph{et al.}~\cite{fisher2014quantum} and enables secure evaluation of non-Clifford gates on encrypted quantum data without revealing the plaintext.}

\label{fig:non-clifford-qhe}
\end{figure}

QHE provides a foundational tool for secure delegated quantum computation by separating data confidentiality from circuit privacy. In our work, QHE is leveraged to handle public circuit layers and enable secure transitions between encrypted states and BQC-protected subcircuits. The ability to track QOTP keys across Clifford layers, and to securely process non-Clifford layers using auxiliary qubits, is critical for the seamless integration of QHE with blind quantum computation frameworks.

\section{Framework for Partial Blind Quantum Computation}
\label{sec:partialbqc}

In this section, we present a framework for applying UBQC to selected subcircuits within a larger quantum computation. A central feature of this approach is how BQC-protected regions interact with the rest of the circuit, which determines both the security properties and implementation feasibility. We classify these interactions into three types based on information flow between protected and unprotected regions:

\begin{itemize}
\item \textbf{Type 0: Connection from unprotected region to BQC-protected subcircuits.} This is a simple case where the output of an unprotected circuit is used as input to a BQC circuit.

\item \textbf{Type 1: Connection from BQC-protected subcircuits to QHE-encrypted region
.} The output qubits from a BQC circuit, typically produced at the final stage, are encrypted via bit-flip and phase-flip operations, following the structure of the QOTP. The client retains the corresponding classical encryption keys. To interface these encrypted output qubits with unprotected circuits, we employ a QHE scheme based on QOTP \cite{fisher2014quantum}. Since both encryption schemes are structurally compatible, BQC output qubits can be interpreted as QHE ciphertexts.

\item \textbf{Type 2: Connection from QHE-encrypted region to BQC-protected subcircuits.} This case arises when QHE-encrypted qubits, originating from unprotected circuits, are to be reused as input to a BQC-protected computation. A naive approach would be to decrypt the qubits before passing them to the BQC circuit, but this would expose plaintext quantum states to the untrusted server. To prevent this, we instead blind the entire composite circuit. Let $QC_D$ denote the decryption circuit for the QHE ciphertext, and $QC_O$ the target computation. While Type~0 applies $BQC(QC_O)$, here we apply $BQC(QC_D||QC_O)$, ensuring that decryption and execution occur entirely within the BQC framework, preserving blindness throughout.
\end{itemize}

\begin{figure}
\centering
\includegraphics[width=0.9\linewidth]{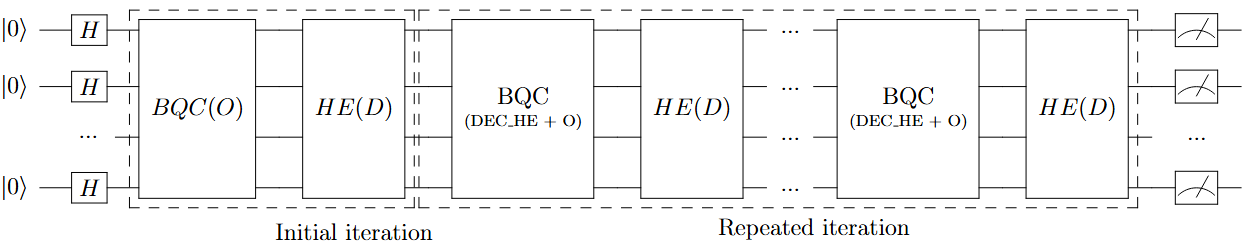}
\caption{Circuit diagram demonstrating the application of the proposed BQC framework to Grover's algorithm.}
\label{fig:grover_example}
\end{figure}

To demonstrate the applicability of our framework, we use Grover’s algorithm as a representative example. Fig.~\ref{fig:grover_example} illustrates how each connection type is instantiated within the circuit structure.

In Grover's algorithm, BQC is applied to the oracle component, while the diffusion transformation is left unprotected, as it does not reveal any sensitive information about the target function $f$. The corresponding connection types within the circuit are as follows:

\begin{itemize}
  \item \textbf{Type 0:} Connection from the initial state preparation ($H^{\otimes n}|0\rangle^{\otimes n}$) to the first BQC-protected oracle $BQC(O)$.
\item \textbf{Type 1:} Connection from the BQC-protected oracle to the unprotected diffusion operator $HE(D)$ within each Grover iteration.
\item \textbf{Type 2:} Connection from the output of the diffusion operator to the input of the next BQC-protected oracle.
\end{itemize}

While Grover’s algorithm is used here as a representative example, the proposed method constitutes a general framework applicable to arbitrary quantum circuits.
It enables flexible application of BQC to selected regions—either individually or in combination—depending on the desired level of security.
This selective control opens up new opportunities for integrating BQC into complex quantum algorithms.

In the following subsections, we detail each connection type, outlining their operational roles and the associated technical considerations.
We note that Type~0 can be regarded as a special case of Type~2, in which all QOTP key bits are zero. Accordingly, we focus our detailed discussion on Type~1 and Type~2.

\subsection{Framework Type 1: UBQC Output Forwarding to QHE Subcircuits}

%\subsection{Framework Type 1: UBQC Output Integration as QHE Input}

In UBQC protocol(BFK09), the output qubits at the final stage of the circuit are encrypted as follows:\\

\[
\ket{\psi^{'}_{n,y}}=X^{S^X_{n,y}}Z^{S^Z_{n,y}} \ket{\psi_{n,y}},
\]

where $\ket{\psi_{n,y}}$ is the desired plaintext result and $(S^X_{n,y}, S^Z_{n,y})$ are classical encryption keys held by the client.
This form corresponds to the QOTP, and is also structurally analogous to QHE \cite{fisher2014quantum}.

Using this QOTP-based encryption, the server can homomorphically evaluate the quantum circuit delegated by the client by applying gates directly to the encrypted qubits, while the client classically tracks and updates the encryption keys.

For Clifford and CNOT gates, the encrypted form is preserved; for non-Clifford gates, such as $T$ gate, auxiliary qubits and additional gate operations (e.g., $P$ and $Z$ gates) are required to maintain the encrypted structure.

For non-Clifford gates, such as the $T$ gate, the client prepares an auxiliary qubit and engages in an interactive, gadget-based protocol involving CNOT gates, measurements, and $P$-gates, enabling the server to homomorphically evaluate the target gate without learning any information about the encrypted data.
The detailed evaluation process, including client-side key updates and auxiliary state preparation, is described in Section~\ref{sec:preliminaries-QHE} (see Fig.~\ref{fig:non-clifford-qhe}).
These transformations can be precompiled into the circuit layout and reused across iterations. For instance, in Grover’s algorithm, the diffusion transformation maintains a fixed structure across repetitions, making it compatible with this QHE-based treatment.

\paragraph{Security Considerations.}
The output of a UBQC subcircuit is inherently protected by the UBQC protocol’s blindness guarantees, which ensure that the server learns nothing about the client’s input, output, or computation beyond the size of the circuit~\cite{broadbent2009universal}. When this encrypted output is used as the input to a QHE-evaluated subcircuit, it retains its security under the QOTP form.

For circuits requiring non-Clifford gate evaluation, the transition from UBQC to QHE includes the preparation of an auxiliary qubit and a single round of classical interaction as specified by the QHE protocol~\cite{fisher2014quantum}. This process does not compromise privacy: the security of the data is preserved by the QHE scheme, which ensures that the quantum server cannot extract any information from the encrypted input or during the interactive evaluation of non-Clifford operations.

As a result, the composition of UBQC and QHE maintains end-to-end confidentiality, with each subprotocol enforcing its own security model at the interface. This modular approach allows the hybrid framework to securely delegate computations that span both interactive and non-interactive quantum components, without introducing new avenues for information leakage.

\subsection{Framework Type 2: Injecting QHE Ciphertexts into UBQC}

As a quantum circuit transitions from a public region, evaluated by the server using QHE, to a privacy-sensitive subcircuit, it becomes necessary to introduce client-side blindness.
At this point, QHE-encrypted qubits must be injected into a UBQC subcircuit without exposing their plaintext states.
A naive solution would be to decrypt the ciphertext before delegation, but this compromises security against an untrusted server.

Instead, we adopt a modified version of the [BFK09] protocol that supports encrypted inputs. A brief overview of this modification is provided below, while a formal description and security analysis are presented in Section~\ref{sec:bqcqfhe}.

After the initial preparation phase, where the client sends random single-qubit states and the server entangles them into a cluster state, the server additionally entangles the QHE-encrypted qubits $X^a Z^b \ket{\psi}$ with the first column of the cluster using CNOT gates, as shown in Fig.~\ref{fig: QHE-encrypted teleportation}.
the server then measures these encrypted qubits, collapsing the cluster into a rotated and re-encrypted form:
\[
X^{m \oplus a} Z^b R_Z(\theta) \ket{\psi}.
\]

To compensate for the QOTP encryption, the client adjusts the measurement angles as:
\[
\phi'_{0,y} = (-1)^{m \oplus a} \phi_{0,y} + \pi b,
\]
and sets the actual measurement angle as:
\[
\delta_{0,y} = \phi'_{0,y} + \theta_{0,y} + \pi r_{0,y}.
\]

the client and the server then proceed with the interaction and measurement phases of the [BFK09] protocol as usual.
This technique ensures that encrypted output qubits from one stage can be securely re-injected into another BQC computation without ever revealing any underlying plaintext state or circuit information, except for the circuit size.\\

\begin{figure}
\centering
\makebox[\textwidth]{
    \Qcircuit @C=1em @R=.7em {
    \lstick{X^aZ^b\ket{\psi}} & \targ      & \qw  & \meter   & \cw   & \cw & \rstick{m} \\
    \lstick{\ket{+_\theta}}    & \ctrl{-1}        & \qw       & \qw         & \qw    & \qw & \rstick{X^{m\oplus a} Z^b R_Z({\theta})|\psi\rangle} \\
    }
}
\caption{
Teleportation-based injection of a QHE-encrypted qubit into a UBQC subcircuit.
The server performs a CNOT and measures the encrypted qubit, transferring the logical state to the auxiliary qubit in a rotated and re-encrypted form.
This procedure allows the secure initialization of UBQC using QHE ciphertexts without revealing any plaintext data or circuit structure.
%Teleportation to inject QHE-encrypted qubits into a BQC subcircuit.}
}
\label{fig: QHE-encrypted teleportation}
\end{figure}

\begin{figure}
\centering
\includegraphics[width=0.8\linewidth]{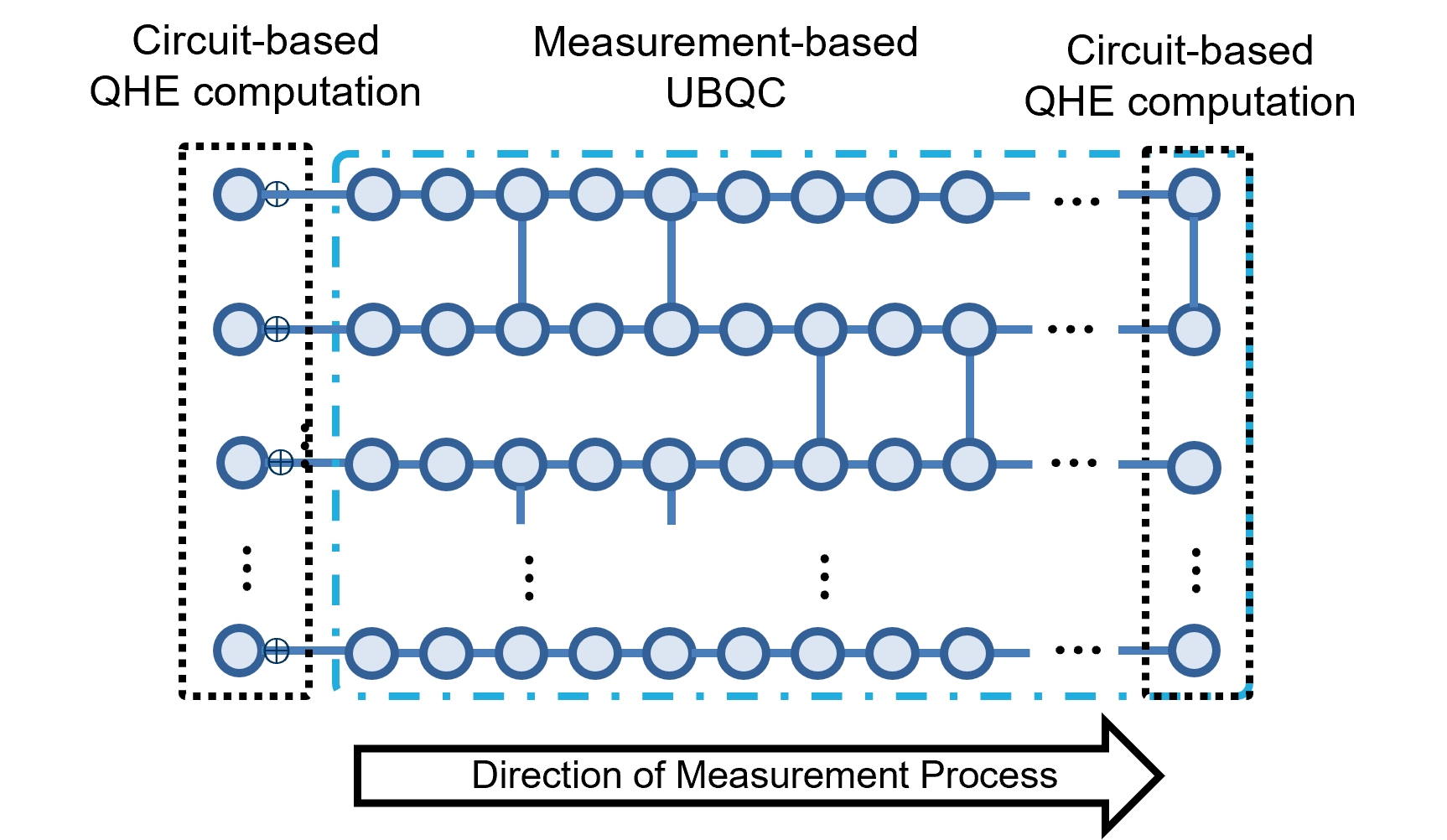}
\caption{
Illustration of a partially blind quantum circuit that combines UBQC and QHE components.
The initial input is QHE-encrypted and injected into a UBQC-protected subcircuit (Type~2), enabling blind evaluation.
The UBQC subcircuit produces a QOTP-encrypted output, which is then forwarded to a QHE-compatible subcircuit (Type~1).
In this hybrid setup, Clifford gates are evaluated non-interactively under QHE, whereas non-Clifford gates require minimal interaction and an auxiliary qubit.
This structure demonstrates how the selective application of blindness and homomorphic encryption can preserve end-to-end confidentiality while improving resource efficiency.
}
%Example circuit combining Type~1 and Type~2: integration of BQC and QHE components with encrypted qubit reuse.
\label{fig:partial_bqc}
\end{figure}

Fig.~\ref{fig:partial_bqc} demonstrates how Type~1 and Type~2 connections can be composed within a single computation to support partially blind quantum circuits. This structure serves as a practical example of hybrid delegation using both UBQC and QHE components.

\section{UBQC Protocol compatible with QHE}
\label{sec:bqcqfhe}

In this section, we present a detailed construction of the modified UBQC protocol that incorporates QHE-encrypted inputs, followed by a formal analysis of its security.

\subsection{QHE-Compatible UBQC Protocol}

The UBQC protocol introduced in [BFK09] enables secure delegated quantum computation by preserving the privacy of both the computation and its results. It provides several variants to accommodate different combinations of quantum inputs and outputs. However, none of the variants provided in [BFK09] directly support QHE-encrypted inputs and outputs, which are essential for our partially blind computation model. This motivates the development of a modified UBQC protocol compatible with QHE.

We refer to this design as the QHE-Compatible UBQC protocol, which accepts QHE-encrypted quantum inputs and produces QHE-encrypted outputs. This construction enables seamless integration of homomorphically encrypted quantum states into the UBQC framework. In this setting, the output remains encrypted, allowing the client to update or propagate the encryption keys without revealing any quantum information to the server.

Our QHE-Compatible UBQC protocol is formally described below.
We adopt the definitions and notation from~\cite{broadbent2009universal}, referring to the client and server as Alice and Bob, respectively, and assuming that Alice has predefined a measurement-based quantum computation (MBQC) pattern on a brickwork state $\mathcal{G}_{n\times m}$ for delegation to Bob.

We consider a quantum state grid of dimension $(n+1) \times m$, where the first column $(x=0, y=1,\dots,m)$ represents a $1 \times m$ input state encrypted under QHE. Each qubit $\ket{\psi_{x,y}}$ in the grid is indexed by its column $x \in \{0,1,\dots,n\}$ and row $y \in \{1,\dots,m\}$, and is processed according to the delegated MBQC pattern.

All qubits in the circuit—except those in the first and last columns—are assigned measurement angles $\phi_{x,y}$ according to the MBQC pattern. Each qubit is associated with a set of X-dependencies $D_{x,y} \subseteq [x-1] \times [m]$ and Z-dependencies $D'_{x,y} \subseteq [x-1] \times [m]$, determined via the flow construction. At measurement time, the actual measurement angle is computed by incorporating Pauli corrections as follows:
\[
\phi'_{x,y} = (-1)^{s^X_{x,y}} \phi_{x,y} + s^Z_{x,y} \pi,
\]
where $s^X_{x,y}$ and $s^Z_{x,y}$ denote the parities of the measurement outcomes within the respective dependency sets.

Based on this setup, Protocol~1 formally defines the QHE-Compatible UBQC protocol.

\begin{tabular}{p{0.9\columnwidth}}
  \Xhline{1pt}
  \noindent\textbf{Protocol 1} QHE-Compatible Universal Blind Quantum Computation\\
  \vspace{0.01cm}

  \noindent\textbf{1. Alice's auxiliary preparation}\\
  \hspace{0.2cm}For each column \( x = 1, \dots, n-1 \):\\
  \hspace{0.3cm}For each row \( y = 1, \dots, m \):\\
  \hspace{0.4cm}Alice prepares \( \Ket{\psi_{x,y}} \in_R \{ \Ket{+_{\theta}} \, | \, \theta = 0, \frac{\pi}{4}, \dots, \frac{7\pi}{4} \} \),\\
  \hspace{0.4cm}and sends the qubits to Bob.\\
  %Where $ \Ket{+_{\theta_{x,y}}} = \frac{1}{\sqrt{2}}(\Ket{0} + e^{i\theta_{x,y}} \Ket{1} )$\\

  \vspace{0.15cm}\\
  \noindent\textbf{2. Alice's input/output preparation}\\
  \hspace{0.4cm}Alice prepares the first column of qubits $\ket{\psi_{0,y}} = X^{a_y} Z^{b_y} \ket{\psi_y}$\\
  \hspace{0.4cm}and the last column of qubits $\ket{\psi_{n,y}} = \ket{+} ( y=1,...,m$).\\
  \hspace{0.4cm}Then, sends the qubits to Bob.

  \vspace{0.15cm}\\
  \vspace{0.01cm}
  \textbf{3. Bob's preparation}\\
  \hspace{0.2cm}Bob entangles all the received qubits based on their respective indices by applying \( CX \) gates between qubits in the first and second columns, where the qubits in the first column act as the target qubits. For all other interactions, \( CZ \) gates are applied, thereby constructing the brickwork state \( \mathcal{G}_{n \times m} \).
  \vspace{0.15cm}\\
  \vspace{0.01cm}
  \textbf{4. Interaction and measurement}\\
  \hspace{0.2cm}For the first column ($x=0, y=1,...,m$), Bob measures\\
  \hspace{0.4cm}in the basis $\{\ket{0}, \ket{1}\}$, and sends the result \( s_{0,y} \in \{0,1\} \) to Alice.\\

  \hspace{0.2cm}For each column \( x = 1, \dots, n-1 \):\\
  \hspace{0.3cm}For each row \( y = 1, \dots, m \):\\
  \hspace{0.4cm}4.1 Alice computes \( \phi'_{x,y} \) with the special case,\\
  \hspace{1.0cm} \( \phi_{1,y}' = (-1)^{s^X_{1,y}\oplus{a_y} } \phi_{1,y} + \pi ( s^Z_{1,y}\oplus b_y ) \).\\
  \hspace{0.4cm}4.2 Alice samples \( r_{x,y} \in_{R} \{0,1\}\) uniformly at random and \\
  \hspace{1.0cm}computes \( \delta_{x,y} = \phi'_{x,y} + \theta_{x,y} + \pi r_{x,y} \).\\

  \hspace{0.4cm}4.3 Alice transmits \( \delta_{x,y} \) to Bob.\\
  \hspace{1.0cm}Bob measures in the basis \( \{\Ket{+_{\delta_{x,y}}}, \Ket{-_{\delta_{x,y}}}\} \).

  \hspace{0.4cm}4.4 Bob sends the result \( s_{x,y} \in \{0,1\} \) to Alice.\\

  \hspace{0.4cm}4.5 If \( r_{x,y} = 1 \), Alice flips \( s_{x,y} \);\\
  \hspace{1.0cm}otherwise, no correction is applied.\\
  \vspace{0.15cm}\\

  \textbf{5. Output}\\
  \hspace{0.2cm}Bob sends to Alice all qubits in the last layer $X^{a'_y} Z^{b'_y} \ket{\psi'_y}$.\\
  \hspace{0.2cm}Alice interprets the measurement parities from the final column as QHE encryption keys, setting $a'_y = s^X_{n,y}$ and $b'_y = s^Z_{n,y}$.
  \vspace{0.15cm}\\
  \Xhline{1pt}
\end{tabular}

\vspace{0.15cm}

\subsection{Security Analysis}

The UBQC protocol presented in~\cite{broadbent2009universal} has been proven to satisfy Universality, Correctness, and Blindness. Since Protocol~1 is a modification of that construction, we focus on verifying that these properties are preserved.

The primary difference between Protocol~1 and the original UBQC protocol lies in the treatment of the input and output columns.
Directly replacing the first column of the UBQC brickwork state with QHE-encrypted input states breaks the blindness property, as it prevents Alice from hiding the measurement angles $\phi'_{1,y}$.
This is because, in the standard UBQC protocol, the randomness in the input state preparation plays a crucial role in concealing measurement dependencies.
To address this issue, Protocol~1 introduces a preliminary teleportation step that embeds the encrypted input into the computation while preserving blindness.
After this integration step, the remaining operations in Protocol~1 follow the same structure as in~\cite{broadbent2009universal}, with only minor adjustments to the Pauli correction calculations.

\begin{theorem}(Universality). The brickwork state $\mc{G}_{n,m}$ is universal for quantum computation. Furthermore, we only require single-qubit measurements under the angles $\{ 0, \pm \pi/4, \pm \pi/2 \}$, and measurements can be done layer-by-layer.
\end{theorem}

\begin{proof} Protocol 1 uses an ($n+1, m$)-dimensional cluster state, where the first column consists of QHE-encrypted input states. Since the actual computation is performed on the ($n, m$) brickwork state using the same MBQC procedure as in [BFK09], Protocol 1 retains the universality property of MBQC.
\end{proof}

\iffalse
\begin{proof}
Protocol 1은 ($n,m$) 디맨션의 brickwork state에 첫 컬럼은 one-time padded로 암호화된  양자 입력으로 하여  총 ($n+1,m$) 디맨션의 큐빗을 사용한다. 실제 연산은  [BFK09]기법과 동일하게 $(n,m)$ brickwork state를 이용하기 때문에  [BFK09]의 universality 증명에 따라 Protocol 1 또한 Universality를 제공한다.
\end{proof}
\fi

\begin{theorem}(Correctness). Assume Alice and Bob follow the steps of \textbf{Protocol 1}. Then the outcome is correct.

\end{theorem}

\begin{proof} In the first column, Bob measures the QHE-encrypted input qubits in the computational basis, teleporting their states to the second column of the brickwork state. During this process, Alice applies Pauli correction using both the QHE encryption keys ($a_y, b_y$) and the standard MBQC correction terms ($s^X_{1,y}, s^Z_{1,y}$), as follows:
\[ \phi_{1,y}' = (-1)^{s^X_{1,y}\oplus{a_y} } \phi_{1,y} + \pi ( s^Z_{1,y}\oplus b_y ) \]
Since the subsequent computation follows the standard UBQC protocol of [BFK09], Protocol 1 preserves correctness.
\end{proof}

\iffalse
\begin{proof}
Bob이 첫번째 컬럼을 관측하여 엘리스의 암호화된 인풋 큐빗들이 ($n,m$) dimension의 brickwork state의 첫번째 컬럼으로 탤레포테이션 된다. 이때 암호화에 사용된 비트 $a_y, b_y$ 값과 MBQC의 Pauli correction를 함께 고려하여 즉

\[ \phi_{1,y}' = (-1)^{s^X_{1,y}\oplus{a_y} } \phi_{1,y} + \pi ( s^Z_{1,y}\oplus b_y ) \]

와 같이 one-time padded 복호화를 error correction 하는 방법으로 수행하여
두번째 컬럼부터는 [BFK09]와 동일하게 Protocol이 진행된다. 따라서 [BFK09]의 protocol의 correctness에 따라 위의 Protocol 1 또한 Correctness를 만족한다.
\end{proof}
\fi

We follow the definition of Blindness from [BFK09].

\begin{definition}
Let $\mathsf{P}$ be a quantum delegated computation on input $X$ and let $L(X)$ be any function of the input.
We say that a quantum delegated computation protocol is \textnormal{blind} while leaking at most $L(X)$ if, on Alice’s input $X$, for any fixed $Y = L(X)$, the following two hold when given $Y$:
\begin{enumerate}
    \item \textit{The distribution of the classical information obtained by Bob in $\mathsf{P}$ is independent of $X$.}
    \item \textit{Given the distribution of classical information described in 1, the state of the quantum system obtained by Bob in $\mathsf{P}$ is fixed and independent of $X$.}
\end{enumerate}
\end{definition}

\begin{theorem}
\textbf{Protocol 1} is blind while leaking at most $(n,m)$.
\end{theorem}

\begin{proof}
Let $(n,m)$ be the dimension of the brickwork state. Note that the universality of the brickwork state guarantees that Bob's creation of the graph state does not reveal anything about the underlying computation (except $n$ and $m$).

Alice's input consists of
\[
(\ket{\psi_{y}} \mid y \in [m]) \quad \text{and} \quad \phi = (\phi_{x,y} \mid x \in [n], y \in [m] )
\]
with the actual measurement angles
\[
\phi' = (\phi_{x,y}' \mid x \in [n], y \in [m] )
\]
being a modification of $\phi$ that depends on previous measurement outcomes. Let the classical information that Bob receives during the protocol be
\[
\delta = (\delta_{x,y} \mid x \in [n], y \in [m])
\]

and let $A$ and $B$ be the two types of quantum systems initially sent from Alice to Bob, where $A$ consists of auxiliary qubits and $B$ consists of input qubits.

To show the independence of Bob's classical information, let
\[
\theta_{x,y}' = \theta_{x,y} + \pi r_{x,y}
\]
(for a uniformly random choice of $\theta_{x,y}$) and define
\[
\theta' = (\theta_{x,y}' \mid x \in [n], y \in [m]).
\]
We have
\[
\delta = \phi' + \theta',
\]
where $\theta'$ is uniformly random (and independent of $\phi$ and $\phi'$), which implies the independence of $\delta$ and $\phi$.

For Bob's quantum information of type $B$, Bob's quantum state on Alice's input $\ket{\psi_y}$ is independent from
\[
\ket{\psi_{0,y}} = X^{a_y} Z^{b_y} \ket{\psi_y},
\]
since it is quantum one-time padded~\cite{ambainis2000, boykin2003}.

For Bob's quantum information of type $A$, fix an arbitrary choice of $\delta$. Because $r_{x,y}$ is uniformly random, for each qubit of $A$, one of the following two cases occurs:

\begin{enumerate}
  \item If $r_{x,y} = 0$, then $\delta_{x,y} = \phi_{x,y}' + \theta_{x,y}'$ and
  \[
  \ket{\psi_{x,y}} = \frac{1}{\sqrt{2}} (\ket{0} + e^{i( \delta_{x,y} - \phi_{x,y}' )}\ket{1}).
  \]
  \item If $r_{x,y} = 1$, then $\delta_{x,y} = \phi_{x,y}' + \theta_{x,y}' + \pi$ and
  \[
  \ket{\psi_{x,y}} = \frac{1}{\sqrt{2}} (\ket{0} - e^{i( \delta_{x,y} - \phi_{x,y}' )}\ket{1}).
  \]
\end{enumerate}

Since $\delta$ is fixed, $\theta'$ depends on $\phi'$ (and thus on $\phi$), but since $r_{x,y}$ is independent of everything else, without knowledge of $r_{x,y}$, $A$ consists of copies of the two-dimensional completely mixed state, which is fixed and independent of $\phi$.
\end{proof}

\begin{theorem} Quantum output of \textbf{Protocol 1}, is one-time padded
\end{theorem}

\begin{proof}
  In Protocol 1, as in the \cite{broadbent2009universal} method, all columns except for the first one do not reveal $s^X$ and $s^Z$ to Bob. This is because they are one-time padded using the random key $r$. Additionally, due to the flow construction, each qubit receives independent Pauli operators. This is equivalent to the random key used in the QOTP. Therefore, the quantum output in the final column remains quantum one-time padded.

\end{proof}

\section{Application and Simulation: 2-Qubit Grover algorithm}

\begin{figure}
\includegraphics[width=1\linewidth]{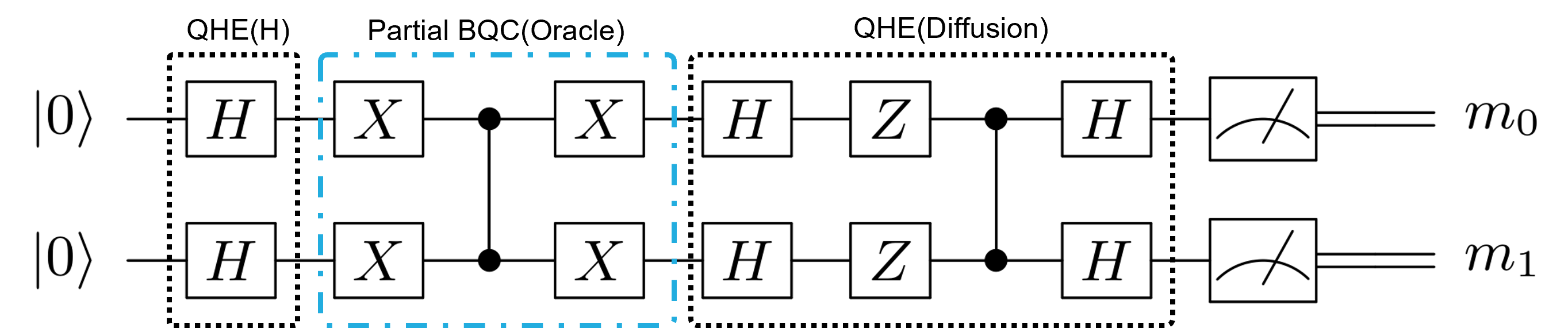}
\caption{Circuit-based 2-Qubit Grover Algorithm}
\label{fig:2_Q_Grover_Circuit}
\end{figure}

\begin{figure}
\centering
\includegraphics[width=0.6\linewidth]{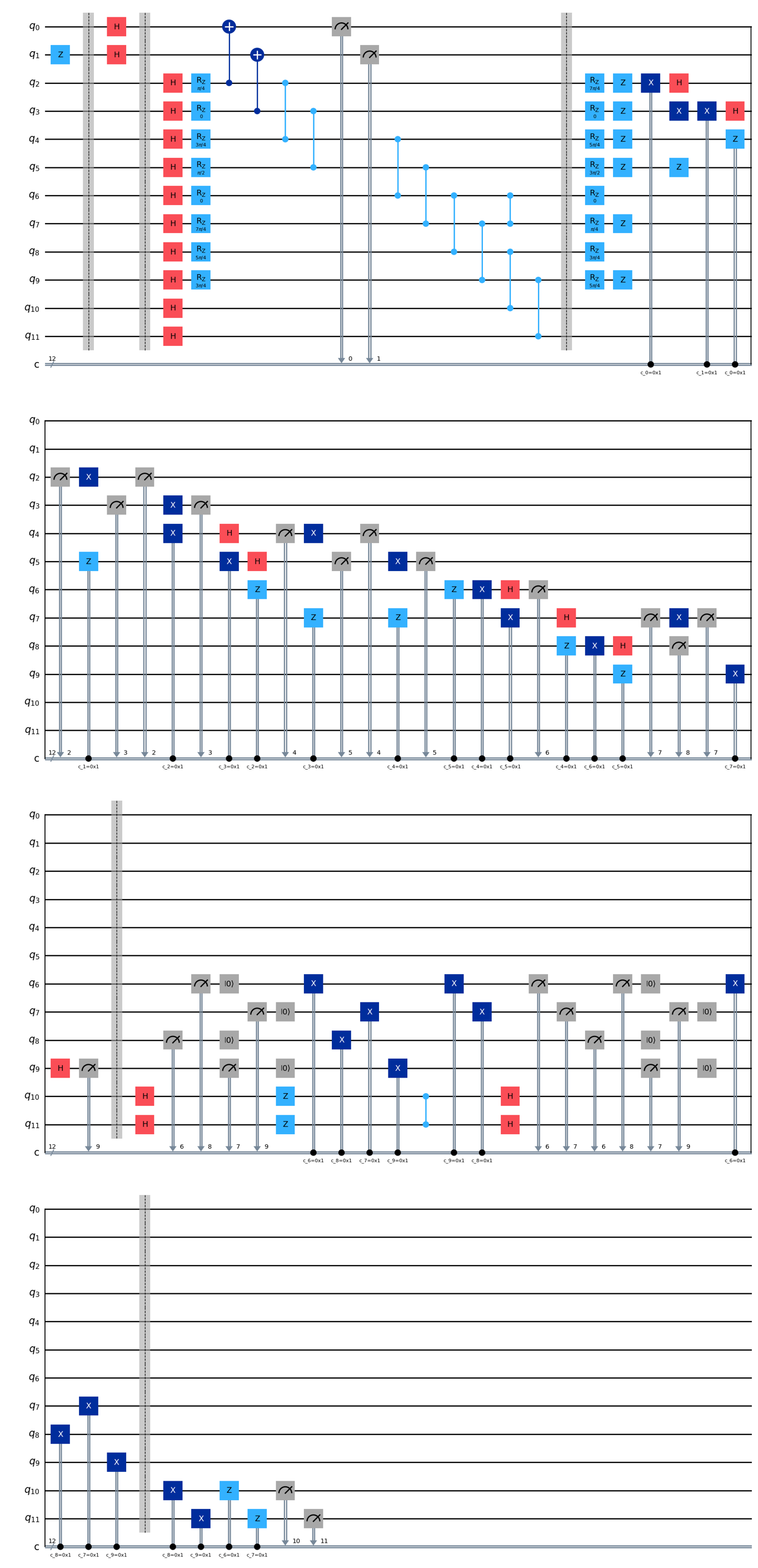}
%\caption{Circuit Simulation of Partial BQC for the 2-Qubit Grover Algorithm}
\caption{
Circuit-level simulation of the proposed Partial Blind Quantum Computation (PBQC) applied to the 2-qubit Grover algorithm. Logical qubits \(q_0\) and \(q_1\) are QHE-encrypted, and the oracle subcircuit—comprising \(X\), \(CZ\), and \(X\) gates—is executed via a proposed QHE-compatible UBQC protocol (Protocol~1) using \(q_2\)–\(q_9\) as the MBQC resource state. After oracle execution, the diffusion operator is applied to QHE-encrypted output qubits \(q_{10}\), \(q_{11}\), with the client updating encryption keys accordingly.
}

\label{fig:Circuit_PBQC}
\end{figure}

This section demonstrates the application of the Partial Blind Quantum Computation framework to a 2-qubit Grover algorithm.
We implement the algorithm using Qiskit~\cite{QiskitGithub, javadiabhari2024}, IBM's open-source software development kit for quantum computing based on the circuit model, and present simulation results to validate the proposed method.

Although Qiskit is a circuit-based quantum computing platform and does not natively support MBQC, recent work has demonstrated that MBQC-based protocols such as UBQC can be implemented within Qiskit. For implementation details, we refer the reader to~\cite{Lee2025}.

The 2-qubit Grover algorithm aims to identify an input \(x\) such that \(f(x) = 1\), where the oracle function \(f\) is considered to encode sensitive information.
To accommodate this, the algorithm is divided into two components: the oracle, which is protected under the Partial BQC framework via QHE-compatible UBQC; and the remaining operations—namely, initialization and diffusion—which are executed using QHE.

The simulation proceeds in the following stages:

\begin{enumerate}
    \item \textbf{Initialization:} \\The client encrypts the initial state \(\ket{00}\) using QHE and sends it to the server.

    \item \textbf{Superposition under QHE:} \\ The server applies the Hadamard (\(H\)) gates to the QHE-encrypted qubits. The client updates the corresponding encryption keys to reflect the transformation.

    \item \textbf{Oracle Execution via Partial BQC:} \\The oracle component of the Grover algorithm, comprising \(X\), \(CZ\), and \(X\) gates to mark the target state \(\ket{11}\), is executed using the QHE-compatible UBQC protocol (Protocol~1). In the simulation, the server designates qubits \(q_0\) and \(q_1\) as input qubits, \(q_2\) through \(q_9\) as auxiliary qubits, and \(q_{10}\) and \(q_{11}\) as output qubits.
        The UBQC simulation is implemented on a circuit-based quantum computing platform, following the procedure described in~\cite{Lee2025}.
        To support full-circuit simulation of the 2-qubit Grover algorithm in Qiskit, a general cluster state is employed instead of the standard brickwork pattern.
        As a result, structural information of the oracle subcircuit, such as the locations of two-qubit gates, is revealed to the server, although blindness with respect to single-qubit measurement angles remains preserved.

    \item \textbf{Diffusion Operation:} \\Following the measurement of qubits \(q_8\) and \(q_9\), the output qubits \(q_{10}\) and \(q_{11}\) remain QHE-encrypted. The server then applies the diffusion operator—comprising \(H\), \(Z\), \(CZ\), and \(H\) gates—to these QHE-encrypted qubits. The client subsequently updates the corresponding encryption keys to reflect the transformation.
\end{enumerate}

A key challenge during the diffusion step is that Qiskit does not permit direct access to classical registers at runtime, which complicates the update of QHE encryption keys. To address this limitation, we implemented a workaround by storing and updating QHE encryption keys using quantum registers instead of classical ones.

In this setup, the encryption keys for the \(X\) and \(Z\) operations are maintained as follows: the \(X\)-gate keys for \(q_{10}\) and \(q_{11}\) are stored in the registers associated with \(q_8\) and \(q_9\), respectively, while the corresponding \(Z\)-gate keys are stored in the registers associated with \(q_6\) and \(q_7\). Although this key update process is adapted for simulation purposes, it is essential for demonstrating the correctness of the computation.

By integrating Partial BQC with QHE and adapting key management for Qiskit-based simulation, we successfully verified the correct execution of the 2-qubit Grover algorithm. The simulation results confirm that the combined approach correctly implements the desired quantum search while preserving the confidentiality of the oracle.

\subsection{Resource Analysis}

To evaluate the resource efficiency of the proposed PBQC framework, we analyze and compare its complexity with that of the standard UBQC protocol and with other state-of-the-art protocols~\cite{broadbent2009universal, morimae2013blind, Mantri2017, zhang2020succinct, poshtvan2025sbqc}. The comparison is presented in both general terms and, specifically, for the 2-qubit Grover search.

\subsubsection{General Case}

We consider a generic quantum circuit acting on \(n\) qubits whose overall depth decomposes as \(d = s + p\), where \(s\) denotes the depth of the privacy-sensitive subroutine (e.g., an oracle in a Grover search) and \(p\) denotes the depth of the public subroutine (e.g., state initialization, diffusion in a Grover search, QFT in a Shor's Algorithm).

In the original UBQC protocol~\cite{broadbent2009universal}, the entire circuit is delegated to the server, which must prepare a universal brickwork state of size $\mathcal{O}(n \cdot d)$ and carry out $\mathcal{O}(d)$ sequential measurement layers. PBQC, in contrast, delegates \emph{only} the sensitive portion.
This reduces the delegated brickwork to $\mathcal{O}(n \cdot s)$ qubits and $\mathcal{O}(s)$ measurement layers, while the public portion of depth \(p\) is executed via quantum homomorphic encryption (QHE).

In relation to prior work, recent BQC protocols can be categorized along two complementary axes: those that remove the quantum requirements on the client and those that aim to cut the quantum overhead on the server.
The \emph{Flow--Ambiguity} protocol of Mantri~\emph{et~al.}~\cite{Mantri2017} typifies the former. 
A completely classical client drives the measurement process on a universal cluster state prepared by the server, while concealing the flow of computation through carefully designed flow ambiguity, thereby eliminating client-side quantum hardware and limiting information leakage to the server.
This convenience, however, provides no savings for the server, the server must still create a universal graph state (and even embeds extra structure to hide the measurement flow), so the size of state and measurement depth remain asymptotically the same as in UBQC.

A similar story holds for measurement-only variants such as the Morimae and Fujii protocol~\cite{morimae2013blind}.
These schemes shift \emph{all} state-preparation and entangling operations to the server and the client needs only a single-qubit measurement device; nonetheless the server still has to prepare a resource of $\mathcal{O}(n\!\cdot\! d)$ qubits and perform $d$ sequential measurement layers.

Zhang \emph{et\,al.}~\cite{zhang2020succinct} propose a ``\emph{succinct}'' variant of BQC in which the client prepares a fixed number $\operatorname{poly}(\kappa)$ of $\kappa$-qubit \emph{gadget} states prior to the computation and can thereafter remain entirely classical.
While this design renders the online phase fully classical, it shifts a significant quantum burden to the offline phase: the client must maintain $O(\kappa)$-qubit quantum memory and prepare large, circuit-independent gadgets whose total size is $O(\operatorname{poly}(\kappa)\cdot \kappa)$.
The cost on the server side is even greater: during an offline ``q-factory’’ phase, the server must generate and store $\kappa$ copies of every logical qubit, amounting to $O(\kappa n d)$ qubits, before these gadgets are distilled into a brickwork state.
In summary, Succinct BQC trades quantum interaction for heavy upfront gadget preparation.
Although it decouples the client's quantum workload from the circuit size—offering potential scalability in large computations-it does not necessarily reduce the quantum memory requirements compared to conventional BQC.

PBQC takes the opposite stance: by confining blindness to the genuinely private subcircuit instead of applying it uniformly to the whole computation, it avoids constructing a large universal cluster in the first place.
Whenever $(s \ll p)$, this yields an immediate asymptotic reduction in both the number of qubits and measurement depth on both server and client sides.

A very recent proposal, SBQC~\cite{poshtvan2025sbqc}, realize \emph{selective blindness}: the client specifies a public set of
$m$ candidate circuits and requires blindness only with respect to \emph{which
one of them is executed} (the so‑called 1‑of‑$m$ security).
To achieve this, the client blinds \emph{only those qubits whose measurement
angles differ across the candidates}, sending one extra qubit per such vertex.
The resulting quantum communication scales as
$O\!\bigl(h(m)\bigr)$, where $h(m)=|\!\bigcup_{i<j}\!\textsf{Diff}(U_i,U_j)|\le nd$ is
the number of masked vertices.
The selectivity, however, shifts the load to the server: all $m$ patterns must be
embedded into a single \emph{merger graph} $G_M$, so the server’s peak memory
grows to $O\!\bigl(g(m)\bigr)$ with
$g(m)=|V(G_M)|$ and $nd \le g(m) \le mnd$—a linear blow-up in the worst case.

%In contrast, PBQC never replicates the public circuit; it runs the $p$‑depth portion in the native gate model and builds a brickwork of size $\Theta(ns)$ for the private part only, simultaneously lowering \emph{both} client communication and server memory whenever $s\!\ll\! p$.

\begin{table}[t]
\caption{Asymptotic resource comparison for an \(n\)-qubit circuit of depth \(d=s+p\)
         (\(s\): privacy‑sensitive depth, \(p\): public depth).
         \(\tau\) denotes the number of non‑Clifford (\(T\)-type) gates inside the public layer \(p\)—
         one auxiliary magic qubit must be sent per such gate.
         \(N\) denotes the number of qubits in the cluster states in~\cite{Mantri2017}.
         \(\kappa\) is the security parameter used by Succinct~UBQC~\cite{zhang2020succinct}.
         For SBQC~\cite{poshtvan2025sbqc},
         \(h(m)\mathrel{:=}\bigl|\bigcup_{i<j\le m}\textsf{Diff}(U_i,U_j)\bigr|\)
         is the number of qubits that must be masked when the client chooses one out of \(m\) public circuits
         (\(0\le h(m)\le nd\)),
         while \(g(m)\mathrel{:=}|V(G_M)|\) is the size of the merger graph that simultaneously embeds all
         \(m\) circuit patterns on the server side
         (\(nd\le g(m)\le mnd\)), determining the server’s peak memory \(O\!\bigl(g(m)\bigr)\).}
  \label{tab:asymptotic-resources}
  \centering
  \begin{adjustbox}{max width=\columnwidth}
    \begin{tabular}{@{}c|c|c|c|c@{}}
      \toprule
      \textbf{Protocol} & \textbf{Number of qubits}&
      \textbf{Meas.\ Depth} & \textbf{Circuit Depth} & \textbf{Security} \\
       & \textbf{Client}/ \textbf{Communication} / \textbf{Server} &
      &  &\textbf{Model}  \\ \midrule
      \cite{broadbent2009universal}           &  $O\!(1)$ /         $O\!(n d)$ /         $O\!(n d)$                & $O\!(d)$ & --- & IT \\
      \cite{morimae2013blind}                & $O\!(1)$     /         $O\!(n d)$ /        $O\!(n d)$                & $O\!(d)$ & --- & IT \\
      \cite{Mantri2017}                      & $0$          /  $0$  /     $O\!(n d)$                & $O\!(d)$ & --- & IT (non-univ, entropy-bound $1.388N$) \\
      \cite{zhang2020succinct}               & $O\!(\kappa)$ / $O\!(\operatorname{poly}(\kappa)\!\cdot\!\kappa)$ / $O\!(\kappa n d)$ (offline) &  $O\!(d)$ & --- & QROM \\
       & 0                     / 0      / $O\!(nd)$ (online) &  &  &  \\

      \cite{poshtvan2025sbqc}                & $O\!(1)$    /  $O\!\bigl(h(m)\bigr)$   /        $O\!\bigl(g(m)\bigr)$ & $O\!(d)$ & --- & 1-of-$m$ \\
      \textbf{PBQC}                   & $O\!(1)$  /  $O\!(ns)+O\!(\tau)$     /   $O\!(ns)+O\!(\tau)$                & $O\!(s)$ & $O\!(p)$ & IT(only $s$ layer) \\ \bottomrule
    \end{tabular}
  \end{adjustbox}
\end{table}

\subsubsection{2-Qubit Grover Case Study}
As a practical example, we apply PBQC to a 2-qubit Grover algorithm.
Fig.~\ref{fig:2_Q_Grover_Circuit} shows that the circuit has total depth
\(d = 9\), consisting of a privacy‑sensitive oracle of depth \(s = 3\) and
public initialization and diffusion layers of depth \(p = 6\).

Under UBQC, the whole circuit is mapped to a cluster state that requires
\(2d = 18\) cluster‑state qubits and 9 measurement layers~\cite{Lee2025}.
PBQC, in contrast, blinds only the oracle by means of its QHE‑compatible
protocol.  As a result, the blinded part needs just 12 cluster‑state qubits
and 5 measurement layers, while the remaining public portion is executed
natively with circuit depth~6.

\begin{table}[h]
\centering
\caption{Resource comparison for the 2-qubit Grover algorithm under UBQC and PBQC.}
\label{tab:grover-resource}
\resizebox{\linewidth}{!}{%
\begin{tabular}{|c|c|c|c|}
\hline
\textbf{Comparison (2-qubit Grover)} & \textbf{Number of qubits} & \textbf{Measurement depth} & \textbf{Circuit depth} \\
\hline
UBQC~\cite{Lee2025} & 18 & 9 & N/A \\
PBQC (our work) & 12 & 5 & 6 \\
\hline
\end{tabular}
}
\end{table}

These results show that PBQC cuts the cluster‑state size by one‑third
(18 → 12) and the measurement depth by almost one‑half (9 → 5), illustrating
the benefit of limiting blindness to the truly sensitive subroutine.

\section{Conclusion and Future Work}
\label{sec:conclusion}

This paper presents a novel framework PBQC, which selectively applies blindness only to the privacy-sensitive portions of a quantum circuit, instead of delegating the entire computation as in traditional UBQC.

Within PBQC, the selected blind subcircuits are executed using a QHE-compatible version of the UBQC protocol, ensuring that protected regions remain hidden from the server while allowing seamless interaction with homomorphically processed public components. By confining blindness to critical subcircuits, our method achieves more efficient quantum computation while preserving the confidentiality of sensitive data.

We defined three types of connections between BQC-protected and unprotected regions, each tailored to support secure quantum data flow with minimal overhead. These connection types formalize how encrypted outputs from one segment can be reused or fed into another, ensuring consistent blindness guarantees. To enable these interactions, we employed QHE, which allows data to move securely between protected and non-protected subcircuits—even in the presence of untrusted servers—without exposing plaintext quantum states.

The effectiveness of the proposed approach was validated through a concrete implementation using the 2-qubit Grover algorithm, where only the oracle subcircuit required blindness. This example demonstrates that our PBQC framework supports selective application of UBQC with component-level granularity, significantly reducing quantum resource overhead while maintaining required privacy guarantees.

In summary, this work presents an efficient and modular framework for selectively applying UBQC to privacy-sensitive components within quantum circuits. By reducing quantum resource overhead and supporting seamless integration into hybrid quantum architectures, the proposed PBQC model improves both practicality and scalability. Moreover, this approach opens promising avenues for future research on adaptive, circuit-aware security mechanisms in quantum computing, with implications for both theoretical advancement and secure quantum protocol deployment.

\section*{Acknowledgment}

This work was supported by Electronics and Telecommunications Research Institute(ETRI) grant funded by the Korean government [25ZS1320, Research on Quantum-Based New Cryptographic System for Ensuring Perfect Data Privacy]

 \bibliographystyle{alpha}
\bibliography{bibtex/ref.bib}

%\input{appendix.tex}

% that's all folks
\end{document}